\documentclass{raa}
\usepackage{graphicx,times}             
\usepackage{natbib}
\bibpunct{(}{)}{;}{a}{}{,}

\usepackage[colorlinks = true, linkcolor = green, anchorcolor = red, citecolor = blue, filecolor = red, pagecolor = red, urlcolor = red]{hyperref}

\begin{document}
  \title{Discovery of Four New Clusters in the Cygnus Cloud}
  
   \volnopage{Vol.0 (20xx) No.0, 000--000}      
   \setcounter{page}{1}          

\author{Songmei Qin \inst{1},
          Jing Li \inst{1,3},
          Li Chen \inst{2,4},
          Jing Zhong \inst{2}}

   \institute{Physics and Space Science College, China West Normal University, 1 ShiDa Rd, Nanchong 637000, China; {\it lijing@bao.ac.cn,chenli@shao.ac.cn,jzhong@shao.ac.cn}\\
   \and
   Key Laboratory for Research in Galaxies and Cosmology, Shanghai Astronomical Observatory, Chinese Academy of  Sciences, 80 Nandan Road,
   Shanghai 200030, China; \\
    \and 
    Chinese Academy of Sciences South America Center for Astronomy, National Astronomical Observatories, CAS, Beijing 100012, China;\\
    \and
    School of Astronomy and Space Science, University of Chinese Academy of Sciences, No. 19A, Yuquan Road, Beijing 100049, China\\
    \vs\no
   {\small Received~~20xx month day; accepted~~20xx~~month day}}


  \abstract
   {We report the discovery of four new open clusters (named as QC 1, QC 2, QC 3 and QC 4) in the direction of Cygnus Cloud and select their members based on five astrometric parameters ($l$, $b$, $\varpi$, $\mu_{\alpha}^*$, $\mu_{\delta}$) of {\it Gaia} DR2. We also derive their astrophysical parameters for each new cluster. Structure parameters are generated by fitting the radial density distribution with King's profile. Using solar metallicity, we performed isochrone-fitting on their purified color-magnitude diagrams (CMDs) to achieve the age of the clusters. The known cluster NGC 7062 at adjacent area is chosen to verify our identification process. The estimated distance, reddening and age of NGC 7062 are in good agreement with the literature. 
   }
   
      \authorrunning{S.-M. Qin, J. Li, L. Chen, \& J. Zhong }            
   \titlerunning{Discovery of Four New Clusters}  

   \maketitle

\section{Introduction}
\label{sec:intro}

Open clusters (OCs), lying close to the Galactic plane, have long been the powerful tool to study the structure and evolution of the Galactic disk \citep{1982ApJS...49..425J,2005ApJ...629..825D,2013ApJ...777L...1F}. Majority of stars in the Galactic disk are formed in clusters \citep{2003ARA&A..41...57L}. Although many clusters were dissolved by dynamical disruption processes, the surviving open clusters have large age and distance range and can be relatively accurately dated. The spatial distribution and kinematic properties of OCs provide critical constraints on the overall structure and dynamical evolution of the Galactic disk, meanwhile, their [Fe/H] values serve as excellent tracers of the abundance gradient along the Galactic disk, as well as many other important disk properties, such as the age-metallicity relation(AMR), abundance gradient evolution, etc \citep{2003AJ....125.1397C,2016A&A...591A..37J}.

Most open clusters are located on the Galactic disk. \citet{2013A&A...558A..53K} listed 3006 star clusters, including about 2700 open clusters, most of which were located within 3~kpc of the Sun. By extrapolating the solar vicinity to the whole disc, about ${10^5}$ open clusters are thought to exist \citep{2006A&A...445..545P}. Most recently, making use of {\it Gaia} DR2 alone, \citet{2020A&A...633A..99C} investigated a list of thousands of clusters from the literature, and homogeneously derived kinematic and distance parameters for 1481 confirmed clusters, in the meantime revealed that some of the clusters from the literature are probably not real clusters. By systematically scanning along the Galactic disk with {\it Gaia} DR2, \citet{2020A&A...635A..45C} detected 582 previously unknown open clusters.

Stars in an open cluster form simultaneously with collapsing molecular clouds (see reviews by \cite{2003ARA&A..41...57L, 2010ARA&A..48..431P}), hence sharing various properties. By assuming that open cluster member stars are co-moving, the VPD (vector-point-diagram) will show a clustering of member stars within the more dispersed field star distribution. Moreover, since they are coeval and located at nearly the same distance, the member stars are expected to show a relatively narrow main-sequence feature on the color-magnitude diagram. Reliably identifying cluster member stars is extremely important and completely necessary for the studies of open clusters, however, the actual reliability of member-star selected by using the astrometric and/or photometric criteria greatly depends on the observational precision. The ambitious European Space Agency (ESA) mission {\it Gaia} \footnote{(\url{https://www.cosmos.esa.int/gaia})} implemented an all-sky survey, which has released its Data Release 2 ({\it Gaia} DR2; \citet{ 2016A&A...595A...1G, 2018A&A...616A...1G}) providing precise five astrometric parameters ($l$, $b$, $\varpi$, $\mu_{\alpha}^*$, $\mu_{\delta}$) and three band photometry ($G$, $G_{BP}$ and $G_{RP}$) for more than one billion stars \citep{2018A&A...616A...2L}. This allows us to investigate a large number of open clusters with unprecedented accuracy.

\begin{figure}[h]
   \centering
   \includegraphics[width=\textwidth, angle=0]{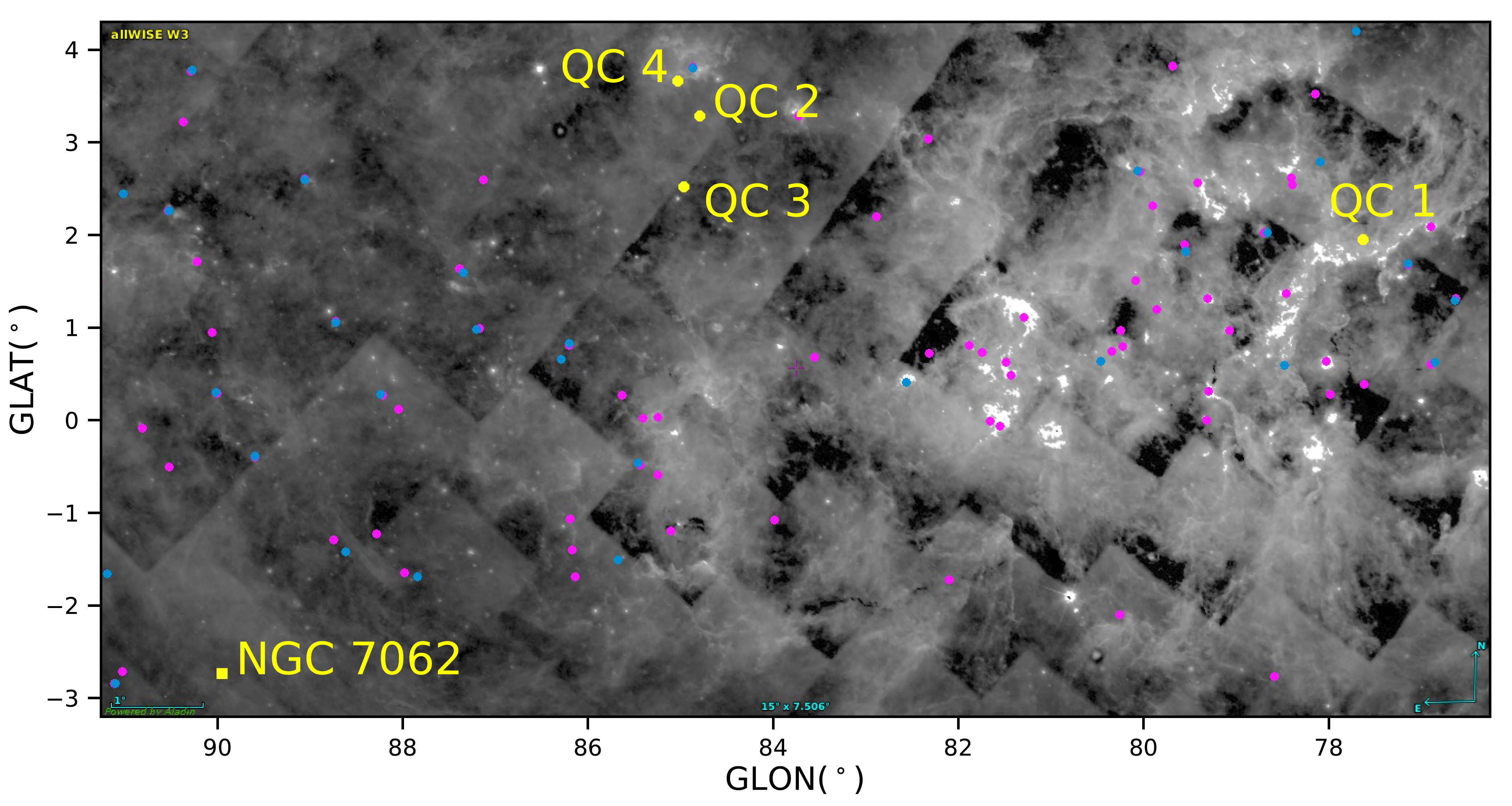}
   \caption{The AllWISE-W3 image taken from the Aladin server within a $15^{\circ} \times 7.5^{\circ}$ size, centered at about $l=83.76^{\circ}$, $b=0.55^{\circ}$. North is up, east is left. It shows the spatial distribution of our newly found OCs (yellow dots: from QC 1 to QC 4) together with the OCs listed by \citet{2013A&A...558A..53K} (pink dots) and \citet{2020A&A...633A..99C} (blue dots) in the Galactic maps. The single yellow square at the bottom-left is the known cluster NGC 7062.}
   \label{cluster}
\end{figure}

The Cygnus Cloud area, usually accepted to be part of the local arm seen tangentially, at a relative proximity of 1-2~kpc with numerous young open clusters and HII regions, is a typical star-forming area. It is also a rather complex area including concentration of dust clouds with heavy and non-uniform absorption and reddening, which leads to the possibility of missing some star clusters in the area. Analyzing the region of Cygnus Nebula Cloud with {\it Gaia} DR2 data, we have noticed the existence of four clusters unreported in the literature as to our best knowledge. Figure~\ref{cluster} is an AllWISE-W3 image around the Cygnus Nebula, in a $15^{\circ} \times 7.5^{\circ}$ size, centered at about $l = 83.76^{\circ}$, $b = 0.55^{\circ}$. WISE\footnote{\url{https://www.nasa.gov/mission_pages/WISE/main/index.html}} is a NASA Medium Class Explorer mission that scanned the sky methodically in the 3.4 $\mathrm{\mu m}$, 4.6 $\mathrm{\mu m}$, 12 $\mathrm{\mu m}$ and 22 $\mathrm{\mu m}$ mid-infrared bandpasses (W1, W2, W3 and W4) in 2010 and 2011 \citep{2010AJ....140.1868W} and the AllWISE program extends the work of the successful WISE by combining data from the cryogenic and post-cryogenic survey phases to form the most comprehensive view of the mid-infrared sky currently available. Overlapped on the image we show the spatial distribution of the centers of our newly found OCs, together with the OCs listed by \cite{2013A&A...558A..53K} and \cite{2020A&A...633A..99C} located in that area, including the well-defined cluster NGC 7062 as a comparing object.

In this paper, based on precise {\it Gaia} DR2 astrometry and photometry data, we confirm the true existence of these four new open clusters (hereafter named as QC 1 through QC 4). In Section~\ref{sec:data}, we present the data used in this work and sample reduction process. In Section~\ref{sec:analysis}, we described the detection of the new open clusters, including the identification and characterization of the clusters. A discussion about this work is given in Section~\ref{sec:discussion}.

\section{Data}
\label{sec:data}

When tried digging up Gaia DR2 in nearby molecular clouds, we serendipitously noticed the existence of four new clusters in the Cygnus Cloud. Thanks to {\it Gaia} DR2, for each of the four new candidate clusters and NGC 7062, we retrieved the positions ($l$ and $b$), parallaxes ($\varpi$), proper motions ($\mu_{\alpha}^*$\footnote{Here $\mu_{\alpha}^*=\mu_{\alpha}cos\delta$}, $\mu_{\delta}$), magnitudes in three photometric filters ($G$, $G_{BP}$ and $G_{RP}$), and their associated uncertainties, in an area with a radius of 0.75$^{\circ}$ centered on the approximate central position. The total numbers of retrieved stars in each of the targeted areas are ranging from 1.3$\times 10^{5}$ to 3.5$\times 10^{5}$. The typical proper motion uncertainty respectively goes from 0.20 mas yr$^{-1}$ for $G \approx$ 17~mag, up to 1.2 mas yr$^{-1}$ for $G =$ 20~mag. The corresponding parallax ($\varpi$) uncertainty goes from 0.1~mas at $G \approx$ 17~mag, up to 0.7~mas at $G =$ 20~mag.

\begin{figure*}[h]
   \centering
   \includegraphics[width=\textwidth, angle=0]{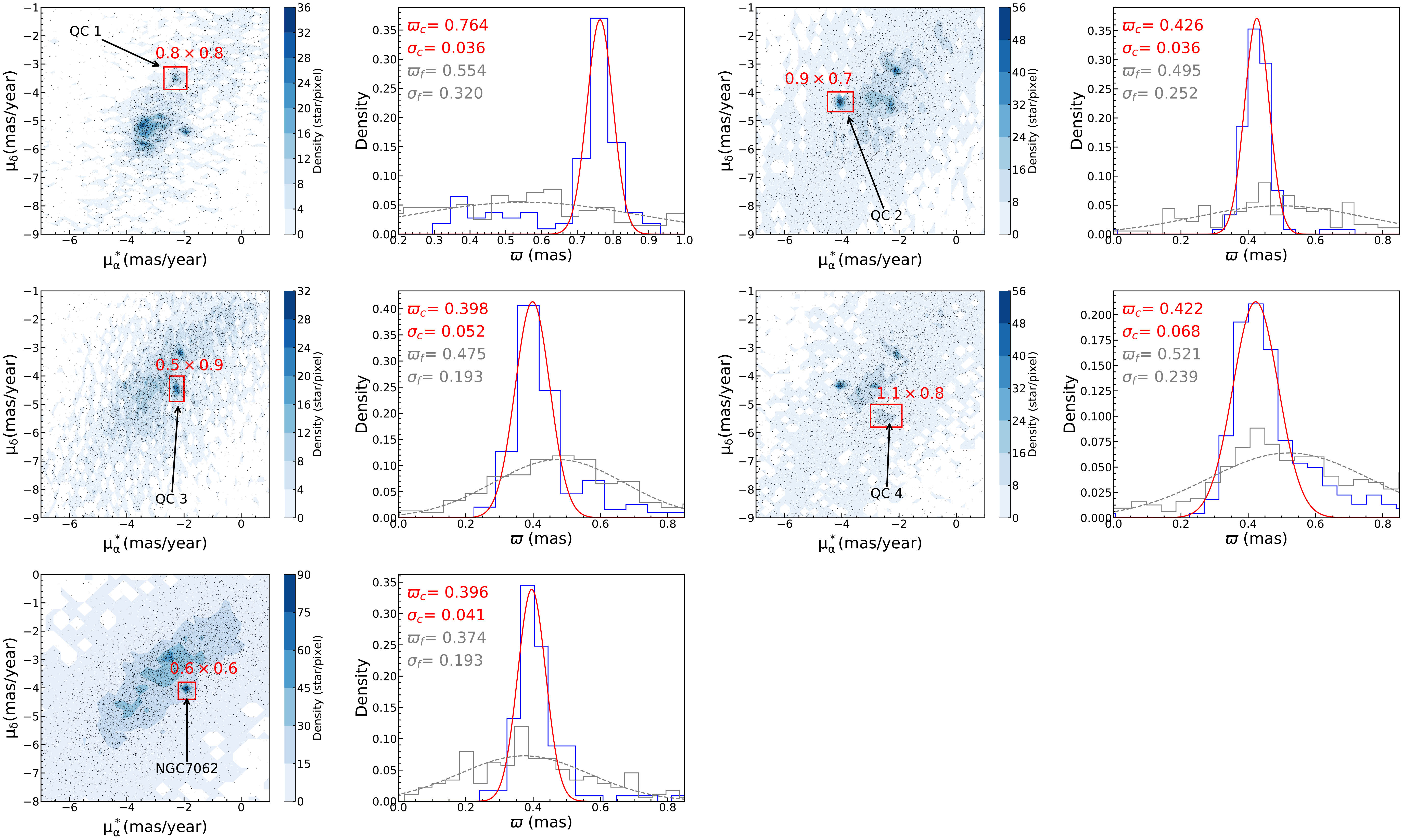}
   \caption{The proper motion distribution and Gaussian fitting of parallax ($\varpi$) for four new clusters and NGC 7062. Proper motion distribution: The over-density of each cluster can be clearly seen and the red rectangles indicate initial proper-motion selecting boxes for each cluster, the color bar on the right represents the number density scale. Gaussian fitting of parallax ($\varpi$): The blue histogram and red line for stars selected by proper motion criterion, the grey histogram and dashed line for the field stars selected randomly from the rest of the stars; and the mean value and standard deviation for candidate members ($\varpi_{c}$, $\sigma_{c}$) and field stars ($\varpi_{f}$, $\sigma_{f}$) are shown at the top left.} 
   \label{PM_PLX}
\end{figure*}

Our target clusters are immersed in a dense star field. It is hard to contrast the cluster members with the background. The most efficient way of hunting out new clusters is to look for the clustering of stars in the velocity space (i.e. the vector-point-diagram), since open cluster members have distinctive movement as compared to the field stars. For highlighting the cluster members in the proper motion distribution diagram, we limited our working sample to magnitude $G < 17$~mag, corresponding to typical astrometric uncertainties of 0.2 mas yr$^{-1}$ in proper motion and 0.1 mas in parallax. In order to further clean our sample, we used re-normalized unit weight error ($RUWE$) $<$ 1.4 as a criterion for selecting sources with ``good'' astrometry, following \cite{LL:LL-124}. This cut seeks to filter out sources with spurious parallaxes or proper motions and retains about 90\% of each cluster sample. Then, we kept the member stars with relative error of parallax $\varpi\_err$/$\varpi < 15\%$. The above selection clearly revealed the over-density features of our new cluster candidates as well as the known target NGC 7062 in the proper motion distribution diagram as shown in Figure~\ref{PM_PLX}.

\section{Analysis and results}
\label{sec:analysis}

\subsection{Identification of the new clusters}
\label{sec:identifying}

To verify the existence of the new clusters by checking the effectiveness of identification process, we took NGC 7062 as an example and carried out the following steps.

We define different proper-motion boxes to contain each of the five co-moving structures (including NGC 7062) to look for members in spatial coordinates (i.e., in $RA$ and $DEC$, $l$ and $b$). The different proper-motion boxes for different clusters are shown in Figure~\ref{PM_PLX}, which were set big enough to isolate the proper motion clump and small enough to mitigate the dispersed field star distribution or other adjacent clumping structures in the proper motion distribution diagram.

Subsequently, for stars in each of the proper-motion boxes, the average proper motion distribution ($\mu_{\alpha}^{*}$, $\sigma_{\mu_{\alpha}^{*}}$; $\mu_{\delta}$, $\sigma_{\mu_{\delta}}$) of candidate members can be acquired by Gaussian-fitting in corresponding directions respectively. In proper motion space, we then used a circular region centered on the expected average proper motion with radius $\sigma_{\mu}$ = $\sqrt{\sigma^{2}_{\mu_{\alpha}^{*}} + \sigma^{2}_{\mu_{\delta}}}$ as the kinematic membership criterion for each candidate cluster.

For the remaining candidate member stars, their parallax distribution exhibits a prominent peak that can be fitted by a Gaussian profile, as the corresponding histogram (blue) and fitting lines (red)  showed in  Figure~\ref{PM_PLX}; the gray histogram and dashed lines refers to the same number of field stars selected randomly in position space from all stars except for those contained in the proper-motion box; the fitted parallax values and standard deviations for cluster candidate members ($\varpi_{c}$, $\sigma_{c}$), field stars ($\varpi_{f}$, $\sigma_{f}$) are marked in the panels. We then excluded stars with parallax ($\varpi$) ranging out of one sigma ($\sigma_{\varpi}$) of the Gaussian distribution peak to further purify the member star samples. 

In summary, in the identification process for the four new OC candidates as well as the comparison cluster NGC 7062 the final sample of member stars for each cluster possess the following features: (i) the proper motion distribution shows an distinct over-density in the corresponding VPD; (ii) the parallax distribution holds an apparent peak, while the projected spatial distribution exhibits certain central concentration  (see the left panels of Figure~\ref{RDP}); and (iii) the selected members in the CMD reveal a relatively narrow main sequence (see in Figure~\ref{isochrone}).

\subsection{Radial density profile}
\label{sec:king}

\begin{figure*}
   \centering
   \includegraphics[width=\textwidth, angle=0]{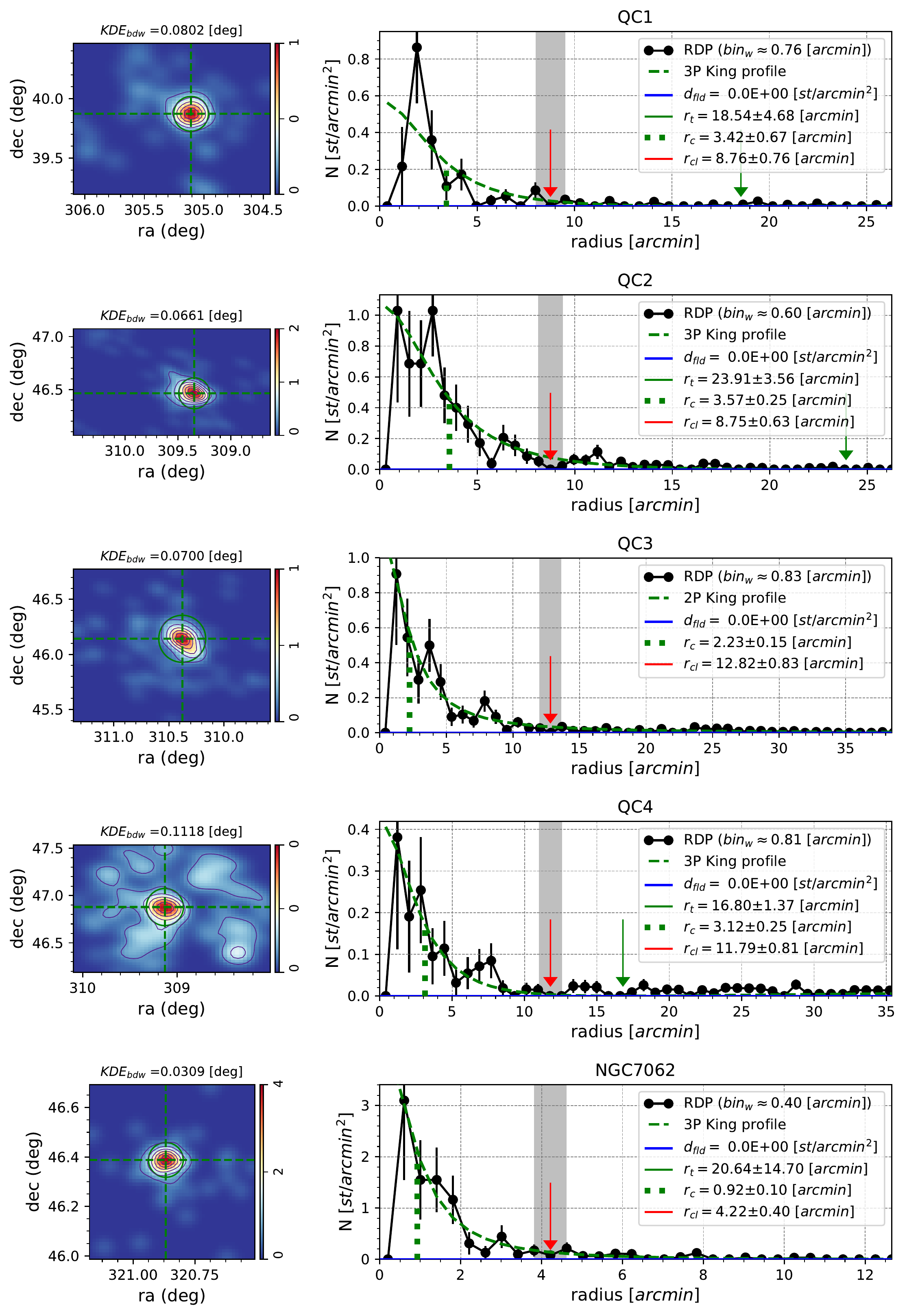}
   \caption{(QC1, QC2, QC3, QC4, NGC 7062) Left panels: center determination via a two-dimensional KDE. Right panels: radial density profile, black dots are the radial density values in each annulus bin; the horizontal blue line is the field density value $d_{\mathrm{field}}$; the red arrow marks the cluster radius with uncertainty region marked as gray shaded area; a King profile fit is indicated with the green dashed curve and a vertical green dotted line denotes the $r_{\mathrm{c}}$ value.}
   \label{RDP}
\end{figure*}

\begin{table*}[h]
\caption{The parameters of newly found OCs and NGC 7062 sorted by increasing $l$. All member stars here are brighter than $G = 17$~mag.}
\label{tab1}
\begin{center}
\begin{tabular}{cccccccccc}
\hline
Name&N&$RA$&$DEC$&$l$&$b$&$\mu_{\alpha}^{*}$& $\mu_{\delta}$&$\varpi$&Dist\\
&&($^{\circ}$)&($^{\circ}$)&($^{\circ}$)&($^{\circ}$)&($mas yr^{-1}$)&($mas yr^{-1}$)&(mas)&pc\\
\hline
QC 1&72&305.106&39.871&77.635&1.932&-2.30$\pm$0.09&-3.48$\pm$0.12&0.766$\pm$0.02&1261\\
QC 2&114&309.349&46.469&84.802&3.278&-4.05$\pm$0.06&-4.32$\pm$0.06&0.424$\pm$0.03&2223\\
QC 3&124&310.381&46.142&84.976&2.513&-2.25$\pm$0.10&-4.46$\pm$0.13&0.400$\pm$0.03&2340\\
QC 4&130&309.117&46.882&85.036&3.654&-2.51$\pm$0.24&-5.43$\pm$0.17&0.424$\pm$0.04&2230\\
NGC 7062&67&320.863&46.384&89.963&-2.744&-1.91$\pm$0.07&-4.05$\pm$0.07&0.395$\pm$0.02&2359\\
NGC 7062$^{\rm a}$&82$^{\rm b}$&320.862&46.385&89.963&-2.743&-1.91$\pm$0.13&-4.05$\pm$0.09&0.398&2343\\
\hline
\end{tabular}
\end{center}
{\textsuperscript{a}{Parameters available from \citet{2020A&A...633A..99C}.}}\\
{\textsuperscript{b}{The number here from \citet{2020A&A...633A..99C} and 82 refers to the cluster star members ($G < 17$~mag, relative error of parallax $<$ 0.15, member probability $>$ 0.6)  of NGC 7062.}}
\end{table*}

\begin{table*}[h]
\caption{The structure parameters of newly found OCs and NGC 7062.}
\begin{center}
\begin{tabular}{cccc}
\hline
Name&$r_{\mathrm{c}}$&$r_{\mathrm{t}}$&$r_{\mathrm{cl}}$\\
&(arcmin)&(arcmin)&(arcmin)\\
\hline
QC 1&3.42&18.54&8.76\\
QC 2&3.57&23.91&8.75\\
QC 3&2.23&--&12.82\\
QC 4&3.12&16.80&11.79\\
NGC 7062&0.92&20.64&4.22\\
\hline
\end{tabular}
\end{center}
\label{tab2}
\end{table*}

We apply ASteCA\footnote{\url{https://asteca.readthedocs.io/en/latest/about.html}} (Automated Stellar Cluster Analysis) code \citep{2015A&A...576A...6P} to determine the  cluster center or maximum spatial density point through a two-dimensional Gaussian kernel density estimator (KDE) fitted on the position space. The spatial distributions of each cluster member stars that satisfy the above selection criteria are also shown in Figure~\ref{RDP} (left panels), which reveal the spatial concentration feature. However, the aggregate structure of some clusters (such as QC 1, QC 4) are relatively sparse, as can be seen from the radial density profiles (RDP) which are also generated with ASteCA.

The ``cluster radius'' $r_{\mathrm{cl}}$ is employed the value where the radial density profile stabilizes around the field density value ($d_{\mathrm{field}}$), which is identical to the ``limiting radius'' $R_{\mathrm{lim}}$ defined in \citet{2005A&A...437..483B}. ASteCA searches the $r_{\mathrm{cl}}$ by using several tolerance thresholds to define when the ``stable'' condition is met. The RDP is obtained by generating concentric square rings via an underlying 2D histogram/grid in the positional space. The first RDP point is calculated by counting the number of stars in the first square ring with a radius of half the bin width, divided by the area of the cell. Repeating the calculus process to get the second RDP point (second square ring, radius of 1.5 bin width), third RDP point (third square ring, radius of 2.5 bin width)$\dots$ Until 75\% of the length of the frame is reached. And the RDP function could mask the bad pixel to correctly avoid empty regions. For each cluster, ASteCA first attempts to three-parameter fit to the RDP and, if that is not possible because of either no convergence or an unrealistic $r_{\mathrm{t}}$, fall back the two-parameter fit \citep{1962AJ.....67..471K,1966AJ.....71..276K,2015A&A...576A...6P}:
\begin{equation}
f(r) = f_{b} + f(0)/[(\frac{1}{\sqrt{1+(\frac{r}{r_{c}})^{2}}} - \frac{1}{\sqrt{1+(\frac{r_{t}}{r_{c}})^{2}}})^{2}]
\end{equation}
or
\begin{equation}
f(r) = f_{b} + f(0)/[1+(r/r_{c})^{2}]
\end{equation}
where $r_{\mathrm{t}}$ means tidal radius, $r_{\mathrm{c}}$ means core radius, and $f(r)$ is the surface number density. The fitting result are shown in Figure~\ref{RDP} (right panels).

For each of the four new clusters and NGC 7062, the number of selected member stars, the cluster central positions (in $RA$ and $DEC$, $l$ and $b$) determined from Gaussian KDE, the mean value and standard deviation of proper motion and parallax found by fitting the Gaussian distribution function are listed in Table~\ref{tab1}, the structure parameters $r_{\mathrm{t}}$, $r_{\mathrm{c}}$, $r_{\mathrm{cl}}$ obtained by ASteCA are given in Table~\ref{tab2}.

\subsection{Isochrone fittings}
\label{sec:isochrone-fitting}

\begin{figure*}[h]
   \centering
   \includegraphics[width=\textwidth, angle=0]{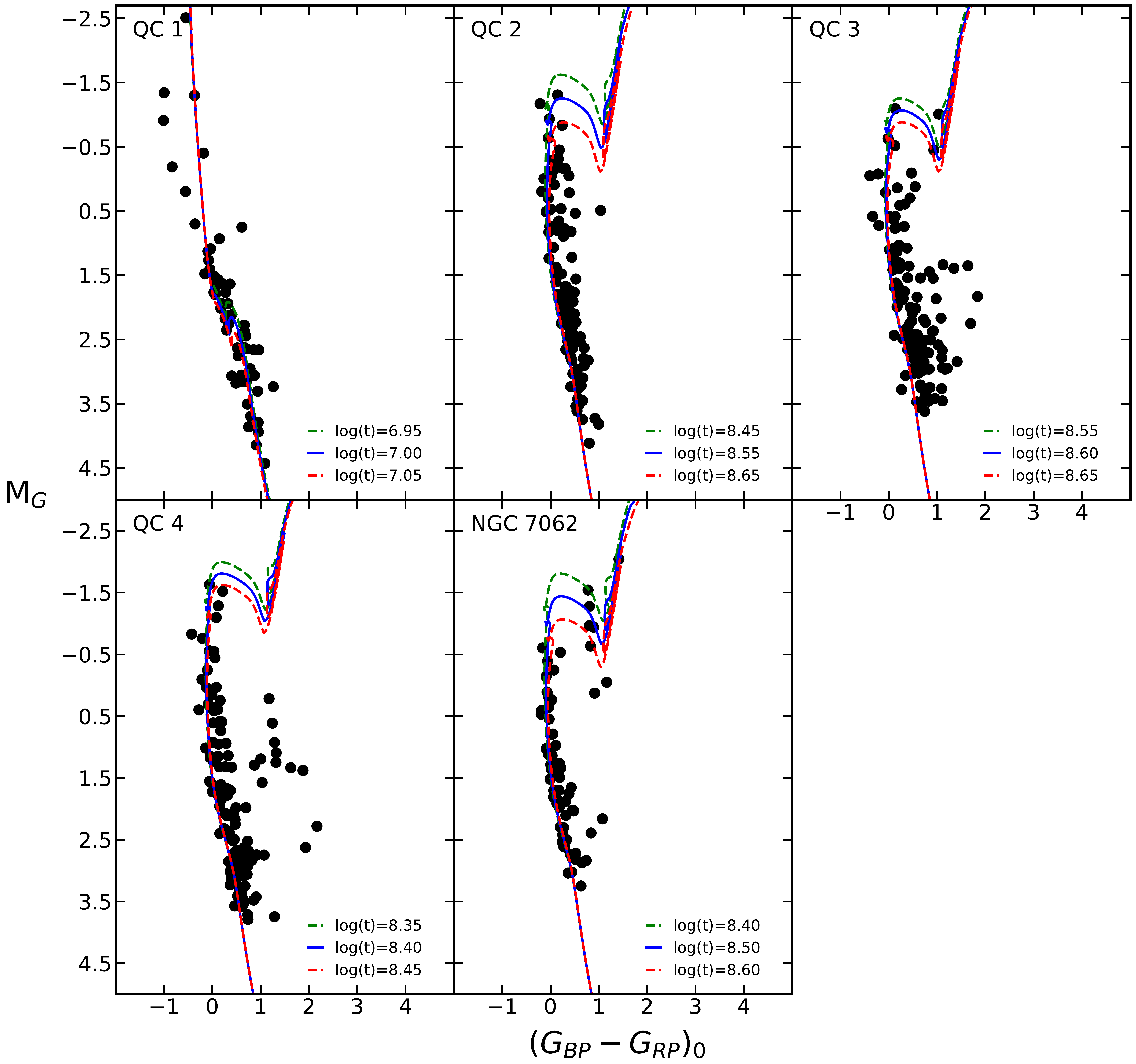}
   \caption{ The CMDs of four new clusters and NGC 7062. Black dots represent member stars from the proper motion and parallax selection. Blue solid lines indicate the best-fitting isochrone while green and red dashed lines denote the possible range of age uncertainties  The three isochrones applied the same solar metallicity ($Z_{\odot}$ = 0.0152). Ages are shown in the panels for each cluster.}  
   \label{isochrone}
\end{figure*}

The color-magnitude diagrams of each cluster were drawn with {\it Gaia} photometry, as shown in Figure~\ref{isochrone}, and the black dots refer to the member stars selected by the criteria in Section~\ref{sec:identifying}. And we perform visual fitting process to make an initial estimate the age parameter of the new open clusters.

To determine the age parameter of the newly discovered clusters and NGC 7062, we retrieved the Padova isochrones \citep{2017ApJ...835...77M} with solar metallicity $Z_{\odot}$ = 0.0152 \citep{2009A&A...498..877C,2011SoPh..268..255C} and {\it Gaia} photometric system \citep{2018A&A...616A...4E} from CMD 3.3\footnote{\url{http://stev.oapd.inaf.it/cgi-bin/cmd/}}, and the logarithm of ages vary in a range of 6.60 to 10.10 with the interval of 0.05. We acquired the reddening values of all the member stars from the Stilism\footnote{\url{https://stilism.obspm.fr/}} 3D dustmaps and the distance from the catalog of 1.33 billion stars in {\it Gaia} DR2 provided by \citet{2018AJ....156...58B}, and we use the ratios $A_{G} = 2.74 \times E(B-V)$ and $E(BP-RP)=1.339 \times E(B-V)$ \citep{2018MNRAS.479L.102C,2019A&A...624A..34Z} in calculating the extinction values. Consequently, we performed an eye-fitting procedure to find the most appropriate isochrone on the observed CMD by shifting in a reasonable range of age parameter.

We carefully inspected the match of the isochrones to the key evolutionary regions, such as the upper main sequence, the turn-off point and the red giant or red clump features on the CMDs. By shifting the isochrones to best fit the CMD of the member stars, we obtained the age of each cluster. Figure~\ref{isochrone} shows the fitting result for four new clusters and NGC 7062, the blue solid line is our adopted best-fitting isochrone, green and red dashed lines denote the possible range of age uncertainties. The astrophysical parameters such as reddening ($E(B-V)$), distance modulus ($m-M$) and age (log(t)) of the four new clusters and NGC 7062 are shown in Table~\ref{tab3} respectively.

\begin{table*}[h]
\caption{The parameters of newly found OCs and NGC 7062.}
\begin{center}
\begin{tabular}{cccc}
\hline
Name&$E(B-V)$&$m-M$&log(t)\\
&mag&mag&\\
\hline
QC 1&0.756$\pm$0.280&12.55&7.00\\
QC 2&0.552$\pm$0.010&13.24&8.55\\
QC 3&0.577$\pm$0.019&13.42&8.60\\
QC 4&0.556$\pm$0.049&13.26&8.40\\
NGC 7062&0.580$\pm$0.051&13.45&8.50\\
NGC 7062&--&--&8.84$^{\rm a}$\\
\hline
\end{tabular}
\end{center}
{\textsuperscript{a}{The age parameter here from \citet{2013A&A...558A..53K}.}}
\label{tab3}
\end{table*}

\section{Discussion}
\label{sec:discussion}

Based on the reliable {\it Gaia} DR2 data, we have identified four new open clusters named QC 1, QC 2, QC 3 and QC 4 in the east part of Cygnus cloud region ($77^{\circ}$ $\le$ $l$ $\le$ $90^{\circ}$ and $-3^{\circ}$ $\le$ $b$ $\le$ $4^{\circ}$). We performed field star decontamination process based on precise {\it Gaia} astrometric data. The selected member stars of these new clusters show apparent proper motion over-density in the corresponding VPD, relatively central concentration in the 3-D spatial distribution and clear main sequence feature on the CMD. For each cluster, we use solar metallicity, the averaged distance values from {\it Gaia} DR2 \citep{2018AJ....156...58B}, and the reddenings from Stilism 3D dustmaps for all individual member stars. Then we performed an eye-fitting procedure to find the most appropriate isochrone on the observed CMD by shifting the age parameter in a reasonable range.

For testing and verifying our member selection and isochrone fitting procedure, we take NGC 7062 as a comparison cluster to go through the same filtering and analysis processes. As for NGC 7062, we analyzed the {\it Gaia} DR2 data in a field of 0.75~deg radius around its center at  $l$ = $89.963^{\circ}$, $b$ = $-2.744^{\circ}$. Finally we obtained 67 member stars brighter than $G =$ 17~mag, of which 47 stars are common to \citet{2020A&A...633A..99C}’s sample (including 82 member stars in total with $G <$ 17~mag,  relative error of parallax $<$ 0.15 and probability $>$ 0.7). The final mean distance(2334~pc) is quite consistent with that from \citet{2020A&A...633A..99C} (Dist = 2343~pc); the mean value of $E(B-V)$ $\approx$ 0.582 for NGC 7062 is larger than the value(0.46) of \citet{1990aadm.conf..141P} and the value(0.32) of \citet{2012NewA...17..720G}, both of which derived from visual isochrone-fitting. The age of NGC 7062 in our work (log(t) = 8.50) is a bit older than the (log(t) = 8.45) of \cite{1990aadm.conf..141P}, but a bit younger than the age(log(t) = 8.84) of \cite{2013A&A...558A..53K} and the age (log(t) = 9.0) of \cite{2012NewA...17..720G}. Our properties for NGC 7062 is in good agreement with that inferred from \citet{2020A&A...633A..99C}, which is also based on {\it Gaia} DR2 data.

In addition, when submitting this paper we find out that in \citet{2019ApJS..245...32L}'s newly published paper, two clusters listed in their catalog (Nr.600, Nr.506) have nearly the same positions for our QC2 and QC3, respectively. In their work results, these two objects are estimated to be very young (4-5 Myr) and categorized as “Class 3” – candidates that need further confirmation. For our four new clusters, more future observations are needed to further investigate their properties. Especially more spectroscopic data for the member stars will be of prime importance to determine the dynamical and chemical nature of these clusters.

\begin{acknowledgements}
We thank the anonymous referee for constructive comments that significantly improve the quality of the paper.This work is supported by National Natural Science Foundation of China (NSFC) under grants 11703019 and China West Normal University grants 17C053,17YC507 and 16E018. L.C acknowledges support from NSFC under grant 11661161016. J.Z would like to acknowledge the NSFC under grant U1731129. 

This work has made use of data from European Space Agency (ESA) mission {\it Gaia} (\url{https://www.cosmos.esa.int/gaia}), processed by the {\it Gaia} Data Processing and Analysis Consortium (DPAC, \url{https://www.cosmos.esa.int/web/gaia/dpac/consortium}). Funding for the DPAC has been provided by national institutions, in particular the institutions participating in the {\it Gaia} Multilateral Agreement.

\end{acknowledgements}

\bibliographystyle{raa}
\bibliography{ms_OC}

\end{document}